\documentstyle[12pt,twoside]{article}
\begin{document}
\begin{center}

{\Large\bf 
CLASSICAL ANALOGOUS OF
QUANTUM\\[5PT]
COSMOLOGICAL PERFECT FLUID MODELS\\[5PT]}
\medskip
 
{\bf A.B. Batista\footnote{e-mail: brasil@cce.ufes.br},
J. C. Fabris\footnote{e-mail: fabris@cce.ufes.br}, S.V.B. Gon\c{c}alves\footnote{e-mail: svbg@if.uff.br} and J. Tossa\footnote{e-mail:
jtoss@syfed.bj.refer.org. Permanent Adress: IMSP - Universit\'e Nationale
du B\'enin, Porto Novo, B\'enin.}}  \medskip

Departamento de F\'{\i}sica, Universidade Federal do Esp\'{\i}rito Santo, 
29060-900, Vit\'oria, Esp\'{\i}rito Santo, Brazil \medskip\\
\medskip

\end{center}
 
\begin{abstract}
Quantization in the mini-superspace of a gravity system coupled to
a perfect fluid, leads to a solvable model which implies singularity
free solutions through the construction of a superposition of the
wavefunctions.
We show that such models are equivalent to a classical system where,
besides the perfect fluid, a repulsive fluid with an equation of state
$p_Q = \rho_Q$ is present.
This leads to speculate on the true nature of this
quantization procedure. A perturbative analysis of the classical system
reveals the condition for the stability of the classical system in terms
of the existence of an anti-gravity phase.

\vspace{0.7cm}

PACS number(s): 04.20.Cv., 04.20.Me
\end{abstract}
 
The existence of an initial singularity is one of the major drawbacks of
the so-called standard cosmological model. It is a general belief that
such problem can be solved through the employement of a quantum theory of
gravity. Indeed, near the singularity sub-Planckian scales are reached and
a classical description of the Universe under this situation is not
appropriate. However, there is no consistent quantum theory of gravity untill
now, and in this sense the problem of the initial singularity remains of
actuality. On the other hand, it is possible to construct a quantum
model for the Universe as a whole, through the Wheeler-de Witt equation,
based in the ADM decomposition of the gravity sector, which leads to a
hamiltonian formulation of general relativity, from which a canonical
quantization procedure can be applied. This gives birth to
quantum cosmology\cite{halliwell,nelson2}.
\par
Quantum cosmology is not free from problems. First, it can be
applied only to geometries where a foliation is possible. Moreover, the
hamiltonian formalism leads to a breakdown of general covariance, and the notion of time is lost
\cite{isham}. There are some recent proposals
by which this notion of time can be recovered. One of these proposals
is based in the coupling of the gravity sector to a perfect fluid.
Using the Schutz's formulation of a perfect fluid \cite{schutz}, a quantization procedure
is possible. The canonical momentum associated with the perfect fluid
appears linearly in the Wheeler-de Witt equation, permiting to rewrite this
equation in the form of a Schr\"odinger equation and a time coordinate
associated with the matter field can be identified.
\par
Solutions based on this approach reveal that a superposition of the
wavefunctions which are solutions of the
resulting Schr\"odinger's equation leads to a singularity-free
Universe\cite{nivaldo,flavio1,demaret,nelson2}. The behaviour of the scale factor may be determined in two
different ways: calculating the expectation value of the scale factor,
in the spirit of the many worlds's interpretation of quantum mechanics;
evaluating the bohmian trajectories in the ontological
formulation of quantum mechanics. The results are essentially the same in
both approachs and
the scale factor display a bounce, the singularity never being reached.
It must be noted that even if the two procedures are technically equivalent,
they are conceptually very different from each other; there are claims that
from the conceptual point of view only the ontological formalism
can be consistently applied to quantum cosmology\cite{nelson}.
\par
The existence of a bounce indicates that there is a repulsive effect, of
quantum origin, when the scale factor approachs the singularity. In this
work we study more in detail such scenario. It is shown that the quantum
scenario can be reproduced exactly by a very simple classical model where
a repulsive fluid is added to the normal perfect fluid. It is surprising
that the repulsive fluid is always the same, given by a stiff matter
equation of state $p_Q = \rho_Q$, independently of the content of the
normal fluid. The existence of this classical analogous of the quantum
model leads us to ask questions on the true nature of the quantification in
this case. Under which conditions
the features of a quantum system can be exactly reproduced by a classical
system? Our analysis is restricted to a perfect fluid coupled to gravity
system, where the notion of time is recovered. But we sketch some considerations on other situations where gravity is coupled to matter
fields.
\par
The existence of a classical analogous of the quantum model allows us
to perform a perturbative analysis establishing under which conditions the
repulsive phase near singularity may be stable or not. In fact, it must
be stressed that a singularity may be avoided through the violation of
the strong energy condition; but, our analysis suggests that the quantum
effects are due to a real anti-gravity phase, which can lead to instabilities
under certain conditions.
\par
In the perfect fluid formulation developed by Schutz, the degrees of
freedom associated with the fluid are given by five scalar potentials
in terms of which the four velocity is written:
\begin{equation}
u_\mu = \frac{1}{\mu}(\phi_{,\nu} + \alpha\beta_{,\nu} + \theta S_{,\nu})
\end{equation}
where $\mu$ is the specific enthalpy. The four velocity is subjected to
the condition
\begin{equation}
\label{c1}
u^\nu u_\nu = 1
\end{equation}
what enables us to express the specific enthalpy in terms of the other
five potentials. The action is then given by
\begin{equation}
\label{a1}
S = \int_Md^4x\sqrt{-g}R + 2\int_{\partial M}d^3x\sqrt{h}h_{ij}K^{ij} +
\int_Md^4x\sqrt{-g}p \quad ,
\end{equation}
where $h_{ij}$ is the metric on the spatial section.
The action (\ref{a1}) is apparently non-covariant because of the pressure
term. But, in fact, the constraints intrinsic to this formalism
permit to recover the
covariance.
\par
We will consider from now on the Robertson-Walker flat geometry
(the spatial section must be compact in order to
be consistent with the boundary conditions)
\begin{equation}
ds^2 = N^2dt^2 - a(t)^2h_{ij}dx^idx^j \quad .
\end{equation}
We assume a barotropic equation of state $p = \alpha\rho$.
Analyzing the conjugate momentum, and eliminating non-physical degrees
of freedom, we can reduce the action (\ref{a1}) to \cite{flavio1}
\begin{equation}
S = \int\biggr\{\dot ap_a + \dot\epsilon p_\epsilon + \dot Sp_S - NH\biggl\}
\end{equation}
where
\begin{equation}
H = - \frac{p_a^2}{24a} - 6ka + p^{\alpha + 1}_\epsilon a^{-3\alpha}e^S \quad .
\end{equation}
There are three cases which will interest us here: $\alpha = - 1$, $\alpha =
\frac{1}{3}$ and
$\alpha = 0$. The first two ones have been studied in \cite{nivaldo,flavio1,nelson2}. Hence,
first we work out in detail the third case, and after we just present
the final results for the first two ones.
\par
Following the Schutz formalism for the description of perfect
fluid, and specializing it to a dust fluid, with $p = 0$, we obtain
the following lagrangian,
\begin{equation}
L = \dot ap_a + \dot\epsilon p_\epsilon - NH
\end{equation}
where $\epsilon$ is the dust variable, with $p_\epsilon$ being its
conjugate momemtum, and $H$ is the hamiltonian 
\begin{equation}
H = - (\frac{p_a^2}{24} + 6ka^2) + ap_\epsilon \quad .
\end{equation}
Classically, this system admits, for the flat case, the well-known dust
solution $a \propto t^\frac{2}{3}$, where $t$ is the proper time, or
equivalently $a \propto \eta^2$, where $\eta$ is the conformal time.
\par
Imposing the quantization
rules
\begin{equation}
p_a \rightarrow - i\frac{\partial}{\partial a} \quad , \quad 
p_\epsilon \rightarrow - i\frac{\partial}{\partial\epsilon}
\end{equation}
and considering that the hamiltonian becomes an operator which acts
on the wavefunction annihilating it,
\begin{equation}
\tilde H\Psi = 0
\end{equation}
we obtain the following partial differential equation governing the
behaviour of the wavefunction:
\begin{equation}
\frac{1}{24}\frac{\partial^2\Psi}{\partial a^2} - ia\frac{\partial\Psi}{\partial\epsilon} = 0 \quad .
\end{equation}
The fact that the conjugate momentum $p_\epsilon$ associated to the
dust fluid variable appears linearly in the hamiltonian, implies
that the Wheeler-de Witt equation in the mini-superspace assumes a
form similar to the Schr\"odinger equation with $\epsilon$ playing the
role of time. Perfoming the redefinition \cite{flavio2}
\begin{equation}
a = \frac{R}{\sqrt{12}} \quad , \quad \epsilon = \frac{t}{\sqrt{12}}
\quad .
\end{equation}
we end up with the equation
\begin{equation}
\label{dust}
\frac{1}{2} \frac{\partial^2\Psi}{\partial R^2} = iR\frac{\partial\Psi}{\partial t} \quad .
\end{equation}
\par
We solve equation (\ref{dust}) using the method of separation of
variables. It leads to the following decomposition of $\Psi(R,t)$:
\begin{equation}
\Psi(R,t) = \xi(R)e^{iEt}
\end{equation}
where $E$ is a (positive) constant, and $\xi(R)$ obeys the equation
\begin{equation}
\label{edust}
\xi'' + 2RE\xi = 0 \quad .
\end{equation}
The prime means derivative with respect to $R$.
The solution for (\ref{edust}) is under the form of Bessel functions:
\begin{equation}
\label{s1}
\xi(R) = \sqrt{R}\biggr(c_1J_{1/3}(\beta R^{3/2}) +
c_2J_{-1/3}(\beta R^{3/2})\biggl) \quad ,
\end{equation}
with $\beta = \sqrt{\frac{8E}{9}}$.
The condition that the hamiltonian operator is self-adjoint, leads to two
possible boundary conditions:
\begin{equation}
\xi(0) = 0 \quad \mbox{or} \quad \xi'(0) = 0 \quad .
\end{equation}
The final results is insensitive to which boundary condition we employ.
Hence, we will work with the first one, but all results are essentially
recovered if the second condition is used.
\par
The general solution is a superposition of (\ref{s1}).
In order to have
analytical expressions, we will use the following superposition:
\begin{equation}
\Psi(R,t) = \sqrt{R}\int_0^\infty \beta^{4/3}e^{-\alpha\beta}
e^{i\frac{9}{8}\beta^2t}J_{1/3}(\beta R^{3/2})d\beta \quad .
\end{equation}
Its solution is \cite{grad1}
\begin{equation}
\label{pack}
\Psi(R,t) = \frac{R}{(2A)^{4/3}}e^\frac{-R^3}{4A}
\end{equation}
where $A = \alpha - i\frac{9}{8}t$.
\par
It may be asked which predictions such model make for the behaviour of
the scale factor. Using the many world's interpretation, this mounts to
evaluate the expectation value of the scale factor. It must be stressed
that essentially the same result is achieved by calculating the
bohmian trajectories \cite{nelson}. The measure employed in the expression
is imposed again by the self-adjoint conditon, and the expression for
the expectation value reads:
\begin{equation}
\sqrt{12}a(t) = <R> = \frac{\int_0^\infty R^2\Psi(R,t)^*\Psi(R,t)dR}{\int_0^\infty
R\Psi(R,t)^*\Psi(R,t)dR} \quad .
\end{equation}
Using (\ref{pack}), the expectation value for the scale factor can
be calculated and the final result is
\begin{equation}
a(t) = \frac{1}{\sqrt{12}}{\frac{2}{\alpha}}^{1/3}\frac{\Gamma(5/3)}{\Gamma(4/3)}
\biggr(\alpha^2 + \frac{81}{64}t^2\biggl)^{1/3} \quad .
\end{equation}
Note that the classical behaviour is recovered for $t \rightarrow \infty$.
But, in general, the quantum model predicts a non-singular model exhibiting
a bounce: when the singularity is approached, quantum effects leads to
a repulsive effect, which leads to a regular transition from a contracting
to an expanding phase.
\par
A dynamical vacuum and a radiative fluid can be analyzed through the same lines
as before.
These problem were treated in \cite{nivaldo,flavio1,nelson2} and
we just present the final results.
\par
The dynamical vacuum is realized through the equation of
state $p = - \rho$. Finding the solutions of the corresponding
Wheeler-de Witt equation, evaluating the expectation value
of the scale factor, it results
\begin{equation}
\label{sdv}
a(T) = \frac{\Gamma(4/3)}{\Gamma(7/6)}\biggr[\frac{64\alpha^2 + 9T^2}{8\alpha}
\biggl]^{1/6} \quad .
\end{equation}
Asymptotically the classical solution for a cosmological constant
is recovered if
we choose, in the classical equations of motion,
the time gauge $N = a^{-3}$ (the identification of the time
coordinates can be justified rigorously \cite{flavio1}). In terms of the proper time, the
solution (\ref{sdv}) can be rewritten as
\begin{equation}
a(t) = \frac{\Gamma(4/3)}{\Gamma(7/6)}\sqrt{8\alpha}
\biggr\{\cosh\biggr[\frac{3}{\sqrt{8\alpha}}
\biggr(\frac{\Gamma(4/3)}{\Gamma(7/6)}\biggl)^3t\biggl]\biggl\}^{1/3} \quad .
\end{equation}
\par
For the radiative case, the wavefunctions
can be also determined through the same procedure and the wave packet constructed. The
scale factor expectation value is given by
\begin{equation}
a(\eta) = \frac{1}{12}\sqrt{\frac{2}{\pi\sigma}}\sqrt{\sigma^2\eta^2
+ (6 - p\eta)^2} \quad ,
\end{equation}
where $\eta$ is the conformal time, $p$ and $\sigma$ being real parameters.
Again, this solution represents a non-singular eternal Universe which coincides asymptotically with the classical radiative solution $a \propto \eta$.
\par
It must be stressed that in all cases, the classical behaviour is recovered
for large values of the proper time. Also, both solutions are singularity free,
with a bounce. Near the bounce repulsive effects appear which, in the 
ontological formulation, are connected with the quantum potential which
corrects the classical equations of motion.
\par
A general feature of the quantum models developped previously
is the appearence of a repulsive phase for small values of the scale factor,
leading to the avoidance of the singularity. In \cite{brasil}, it was
shown that a repulsive gravity single fluid model can lead to
consistent cosmological models if the curvature is negative; however, its
stability is not assured in the absence of ordinary (attractive) matter. Another way
of implementing a repulsive phase in classical cosmology is to consider
two fluids, one that acts attractively, and the other that acts repulsively.
In this case, we may have consistent solutions with flat spatial section.
It is desirable that the repulsive fluid dominates for small values of
the scale factor, whereas the attractive fluid dominates for large values
of the scale factor. Hence, in the general, considering just the flat
spatial section, we may obtain possible consistent models from
\begin{equation}
\label{em}
3\biggr(\frac{\dot a}{a}\biggl)^2 = 8\pi G\biggr(\rho_M - \rho_Q\biggl) =
\frac{C_1}{a^m} - \frac{C_2}{a^n} \quad ,
\end{equation}
where $p_M = \alpha_M\rho_M$, $p_Q = \alpha_Q\rho_Q$,
$m = 3(1 + \alpha_M)$ and $n = 3(1 + \alpha_Q)$. The subscripts $M$ and
$Q$ stand for "normal" matter component and for "quantum" repulsive
component. 
\par
Ordinarilly, normal matter may be a cosmological constant,
dust or a radiative
fluid, corresponding to $\alpha_M = -1, 0, \frac{1}{3}$, respectivelly.
Since it is desirable that the repulsive component dominates at small
values of $a$, then $\alpha_Q > \frac{1}{3}$. We choose then a repulsive
stiff matter fluid $\alpha_Q = 1$, what leads to $n = 6$.
Hence, we will solve the equation (\ref{em}) with $n = 6$ and $m = 0,
3$ and $4$. The solutions are the following:
\begin{eqnarray}
& &\mbox{$\alpha_M = -1$, $\alpha_Q = 1$:} \quad
\label{bs1}
a(t) = \biggr(\frac{C_2}{C_1}\biggl)^{1/6}\cosh^{1/3}3\sqrt{C_1}t \quad ;\\
& &\mbox{$\alpha_M = 0$, $\alpha_Q = 1$:}\quad 
\label{bs2}
a(t) = \biggr(\frac{C_2}{C_1}\biggl)^{1/3}\biggr[\frac{9}{4}\frac{C_1}{C_2}t^2
 + 1\biggl]^{1/3} \quad , \\
& &\mbox{$\alpha_M = 1/3$, $\alpha_Q = 1$:}\quad
\label{bs3}
a(\eta) = \sqrt{\frac{C_2}{C_1}}\biggr[\frac{C_1^2}{C_2}\eta^2 + 1\biggl]^{1/2} \quad .
\end{eqnarray}
\par
The comparison of the above solutions with those obtained through
the construction of a superpositon of
the wavefunctions resulting from the
Wheeler-de Witt equation in the mini-superspace with only
the ordinary perfect fluid, reveals that they are
the same. Hence, wave packets constructed from a quantum
model where, besides the scale factor, there is a perfect fluid matter
degree of freedom (which leads to a time coordinate) are equivalent
to a classical model where gravity is coupled to the same perfect fluid
plus a repulsive fluid with a stiff matter equation of state $p_Q =
\rho_Q$. It is really surprising that the repulsive fluid, in the
classical model, is the
same irrespective of the normal fluid employed in the quantum model.
\par
This equivalence of the quantum model with a classical system different from
that used in quantification process makes us to express our doubts on the
true nature of the quantization scheme for this case. The question
of reproducing the classical equations of motion from the quantum ones appears
already in the ordinary quantum mechanics, and they are expressed in the
so-called Ehrenfest's theorem. According to this theorem, the center of
the wave packet may follow a classical trajectory under certain conditions.
Explicitly, taking the expectation values of the Heisenberg's equations
for the position and momentum operators for a particle
of mass $m$ in a potential
$V(\vec r)$, we find \cite{cohen}
\begin{eqnarray}
<{\dot{\vec r}}> &=& \frac{\vec p}{m} \quad ,\\
<{\dot{\vec p}}> &=& - <\nabla V(\vec r)> \quad .
\end{eqnarray}
These relations coincide with the classical one only if
$<\nabla V(\vec r)> = \nabla V(<\vec r>)$. This happens only for very
special forms of the potential, the harmonic oscillator being an example.
Only in these special cases, we may say that the center of the wave
packet follows a classical trajectory. 
\par
However, the situation discussed here is somehow different from that
analyzed in the Ehrenfest's theorem. In fact the expectation value
of the scale factor of a quantum model derived from gravity and a
perfect fluid of attractive nature is reproduced by a classical
model where another fluid, of repulsive nature, appears. This seems somehow
mysterious.
\par
Some insights into what is happening in this case may come from the
employement of the ontological interpretation of quantum mechanics.
In this case, the problem of time is solved in any situation (not
only when a matter field is present). In fact, the ontological interpretation
predicts that the system follows a real trajectory given by the equations
\begin{equation}
\label{dBB}
p_q = S_{,q} ,
\end{equation}
where the subscript $q$ designates one of the degrees of freedom of the system,
and $S$ is the phase of the wavefunction, which is written as
$\Psi = Re^{iS}$, $R$ and $S$ being real functions.
The equation of motion (\ref{dBB}) is governed not only by
a classical potential $V$ but also by a quantum potential $V_Q = \frac{\nabla^2R}{R}$.
\par
These considerations suggest that the quantum potential has, at least
in the case of the quantization of perfect fluid systems,
a very clear behaviour which
can be classically reproduced by a repulsive stiff matter fluid.
However, we must stress that even in this case
we must find first the wavefunction, through the Wheeler-de Witt equation,
determining than its phase, from which the bohmian
trajectories is computed. In the quantum models studied previously,
it is not
possible, in principle, to identify a classical and quantum potential
in terms of the scale factor from the begining. Moreover, even if this
would be possible, the classical analogous we have determined are completely
independent of the Wheeler-de Witt equation.
\par
Since we have a classical analogous of the bounce models determined through
the Wheeler-de Witt equation, we can investigate if the repulsive effect
leading to the avoidance of the singularity may spoil the stability of the
model. First of all we define what we understand here by instability. A cosmological model is considered unstable if the perturbative variables
diverge when all background quantities are finite. Here, we will consider
a weaker condition:
the model is unstable if the perturbed quantities takes very
large values in comparison with the background quantities, even if they
are not divergent. This is due to the fact that, if this happens, the
hypothesis of homogeneity and isotropy, employed in the definition of
the background, are compromised.
\par
Let us consider our non singular classical system. It can be written
as
\begin{eqnarray}
\label{pe1}
R_{\mu\nu} &=& 8\pi G\biggr[{\stackrel{M}{T}}_{\mu\nu} - \frac{1}{2}g_{\mu\nu}{\stackrel{M}{T}}\biggl]
- 8\pi G\biggr[{\stackrel{Q}{T}}_{\mu\nu} - \frac{1}{2}g_{\mu\nu}{\stackrel{Q}{T}}\biggl] \, , \\
\label{pe2}
{\stackrel{M}{T}^{\mu\nu}}_{;\mu} &=& 0 \, \\
\label{pe3}
{\stackrel{Q}{T}^{\mu\nu}}_{;\mu} &=& 0 \, .
\end{eqnarray}
We perturb these equations in the usual way, introducing the
quantities $g_{\mu\nu} = \stackrel{0}{g}_{\mu\nu} + h_{\mu\nu}$,
$\rho_M = \stackrel{0}{\rho}_M + \delta\rho_M$,
$\rho_Q = \stackrel{0}{\rho}_Q + \delta\rho_Q$. The computation of the
perturbed equations is quite standard\cite{weinberg}, and we just present the final
equations, at linear level:
\begin{eqnarray}
\ddot h + 2\frac{\dot a}{a}\dot h =\frac{1}{\alpha_M - \alpha_Q}
\biggr[- (1 + 3\alpha_M)\biggr(2\frac{\ddot a}{a} + (1 + 3\alpha_Q)
\frac{\dot a^2}{a^2}\biggl)\Delta_M & & \nonumber \\
+ (1 + 3\alpha_Q)\biggr(2\frac{\ddot a}{a} + (1 + 3\alpha_M)
\frac{\dot a^2}{a^2}\biggl)\Delta_Q\biggl] & &\, ,\\
\nonumber\\
\dot\Delta_M + (1 + \alpha_M)\biggr(\Psi - \frac{\dot h}{2}\biggl) &=& 0\, ,\\
\nonumber\\
(1 + \alpha_M)\biggr[\dot\Psi + (2 - 3\alpha_M)\frac{\dot a}{a}\Psi\biggl]
- \frac{n^2}{a^2}\alpha_M\Delta_M &=& 0 \, ,\\
\nonumber \\
\dot\Delta_Q + (1 + \alpha_Q)\biggr(\theta - \frac{\dot h}{2}\biggl) &=& 0\, ,\\
\nonumber\\
(1 + \alpha_Q)\biggr[\dot\theta + (2 - 3\alpha_Q)\frac{\dot a}{a}\theta\biggl]
- \frac{n^2}{a^2}\alpha_Q\Delta_Q &=& 0 \, .
\end{eqnarray}
In these expressions, we have used the following definitions:
$h = \frac{h_{kk}}{a^2}$, $\Delta_M = \frac{\delta\rho_M}{\rho_M}$,
$\Delta_Q = \frac{\delta\rho_q}{\delta\rho}$, $\Psi = \delta u_M^i$,
$\theta = \delta u_Q^i$, where $\delta u_M^i$ and $\delta u_Q^i$ are
the perturbations on the four velocity of the normal and repulsive
fluid respectively.
\par
The perturbed equations presented above do not seem to admit analytical
solutions for the background solutions (\ref{bs1},\ref{bs2},\ref{bs3}).
Hence, we are obliged to perform a numerical integration.
In figures $1$, $2$ and $3$ we display the evolution of density perturbations
for the exotic fluid for the cosmological constant, dust and radiative
cases respectively, in the long wavelength limit $n \rightarrow 0$.
The other perturbed quantities exhibits essentially
the same features.
\par
From the numerical study performed we can expose the following conclusions.
When the cosmological constant is coupled to the repulsive fluid,
the bounce model is unstable. Approaching the minimum of the scale factor
(which in all three cases occurs in the origin),
the perturbations diverge. However, when the dust or radiative fluid
is coupled to the repulsive fluid, the perturbations behave regularly
during all the evolution of the Universe. Hence, these models are stable.
The different behaviours for the cosmological constant and the other cases
may be easily understood. Indeed, in \cite{brasil} a stability study was
performed for the same repulsive fluid (represented there by a free scalar
field); the curvature
was taken to be negative. Instabilities were found in the absence of
ordinary matter. What there is in
common between the analysis made in \cite{brasil} and the present cosmological
constant plus repulsive fluid model is that when the normal fluid decouples from
the repulsive fluid, the repulsive fluid becomes the only one to
contribute to the
right hand side of (\ref{pe1}): it drives an
unstable behaviour. In the other cases (dust and radiation), both fluids
appear in the right hand side of (\ref{pe1}), and hence the perturbations
of the normal and repulsive fluid do not decouple; their interaction
stabilizes the model.
\par
In spite of the fact that we display the results for $n = 0$ only,
for other values of $n$ the features are very similar. For very large
values of $n$, the perturbations exhibit strong oscillations, and
they become divergent near the minimum of the scale factor for the
cosmological constant case. Of course, we have studied the stability
of the classical analogous of the quantum model. But, this study leads
to some insights to what happens in the original framework.
\par
The main point of the present work is that the quantum model, derived from
the Wheeler-de Witt equation for gravity plus perfect fluid through the
Schutz's formalism, has a classical analogous. In this classical analogous
system, the perfect fluid is coupled to another perfect fluid, with a fixed
equation of state $p_Q = \rho_Q$ which appears with a "wrong" sign for the
gravitational coupling. The existence of this classical analogous for
all equations of state of the normal fluid studied in this work, rises doubts about the true
quantum nature of the original quantum cosmological scenario.
It is not clear to us how to solve this doubts for the moment. But, the
existence of a classical system reproducing different quantum models
may indicate that the quantization of a gravity system in the
mini-superspace may be not a real quantization and a more carefull analysis
of this problem is deserved.
\par
In the analysis performed previously, it has been considered a specific
superposition of the solutions of the Wheeler-de Witt equation. Since these
solutions are not square integrable, a superposition of them is in fact
a necessity in those models. It can be argued that other
types of superpositions are possible which may not be in agreement with
the classical analogous treated here; all the richness of the original
quantum model would not be reproduced by the classical model. 
However, the superposition procedure must agree
with physical requirements as, for example, the localization of the wave packet, what is the case of the preceding examples. We may guess that
other possible superpositions, satisfying the same physical requirements,
will lead to essentially the same results.
\par
A perturbative study was performed in the classical model.
It can happens that a bounce model, where the avoidance of the singularity is obtained
through an anti-gravity phase, may not be a stable model. We have verified
that when the normal fluid decouples
from the other perturbed equations, in
such a way that the metric perturbation are coupled to the repulsive fluid
only, the background model is unstable. Otherwise, we can obtain stable
singularity-free models with an anti-gravity phase.
\par
We must stress that the classical analogous model reveals that
the "quantum effects" exhibit an anti-gravity behaviour. That is,
the singularity is avoided with conditions much more stronger than the
simple violation of the strong energy condition, as it happens in many
others singularity-free models \cite{picco}. It must also be emphasized that
all considerations have been done for a perfect fluid quantum model. It should
be important to verify if the correspondance found here remains when
gravity are coupled to matter fields. In \cite{nelson1} the case of
a free scalar field was analyzed. However, a free scalar field is
equivalent to stiff matter. Consequently, the classical analogous
(if still valid for this case)
would contain two kinds of stiff matter, an attractive one and a repulsive
one. Perhaps, the strange behaviour founded in \cite{nelson1}, with
the quantum phase being recovered for large values of the scale factor,
is due to this fact. This specific case deserves to be analyzed.
\vspace{0.5cm}
\newline
{\bf Acknowledgements:}
We have benefited of many enlightfull discussions with
N. Pinto-Neto and Nivaldo Lemos. A.B.B., J.C.F. and S.V.B.G. thank CNPq (Brazil) for financial support.
J.T. thanks CAPES (Brazil) for financial support and the Department of
Physics of the Universidade Federal do Esp\'{\i}rito Santo for hospitality.

\newpage
\centerline{\bf Figure captions}
\vspace{2.0cm}
\noindent
\vspace{0.5cm}
Figure 1: Behaviour of $\Delta_Q(t)$ for $n = 0$ with cosmological
constant.\\
\vspace{0.5cm}
Figure 2: Behaviour of $\Delta_Q(t)$ for $n = 0$ with a dust fluid.\\
\vspace{0.5cm}
Figure 3: Behaviour of $\Delta_Q(t)$ for $n = 0$ with a radiative fluid.\\
\vspace{0.5cm}
\end{document}